\documentclass[aps, pra, showkeys, reprint, nofootinbib, longbibliography]{revtex4-1} 

\usepackage[usenames,dvipsnames]{color} 
\usepackage{amsmath} 
\usepackage{txfonts} 
\usepackage{upgreek} 
\usepackage{todonotes} 
\usepackage{amssymb} 
\usepackage{dsfont} 
\usepackage{braket} 
\usepackage{array} 
\usepackage[unicode,breaklinks]{hyperref}
\hypersetup{
    unicode=true,
    a4paper=true,
    plainpages=false,
    pdftitle={Lattice-depth measurement using multi-pulse atom diffraction outside the weak diffraction limit},
    pdfauthor={B. T. Beswick, S. A. Gardiner, I. G. Hughes},
    pdfsubject={Lattice-depth measurement using multi-pulse atom diffraction outside the weak diffraction limit},
    colorlinks=true,
    linkcolor=blue,
    citecolor=blue,
    filecolor=black,
    urlcolor=blue
}
\hypersetup{colorlinks=true,pdfborder={0 0 0}} 
\usepackage{upgreek} 

\usepackage[caption=false]{subfig} 
\usepackage{graphicx} 
\usepackage[countmax]{subfloat}
\usepackage{import} 
\usepackage{transparent} 
\usepackage[export]{adjustbox}

\definecolor{MyGreen}{rgb}{0,0.5,0} 

\newcommand{\tvect}[2]{%
   \ensuremath{\Bigl(\negthinspace\begin{smallmatrix}#1\\#2\end{smallmatrix}\Bigr)}}

\graphicspath{ {../Figures/} }

\begin{document}

\title{Lattice-depth measurement using multi-pulse atom diffraction in and beyond the weakly diffracting limit}
\author{Benjamin T. Beswick}\email{b.t.beswick@durham.ac.uk} 
\author{Ifan G. Hughes}\email{i.g.hughes@durham.ac.uk}
\author{Simon A. Gardiner}\email{s.a.gardiner@durham.ac.uk}
\affiliation{Joint Quantum Centre (JQC) Durham--Newcastle,
Department of Physics, Durham University, Durham DH1 3LE, United Kingdom}

\date{\today}

\begin{abstract}
Precise knowledge of optical lattice depths is important for a number of areas of atomic physics, most notably in quantum simulation, atom interferometry and for the accurate determination 
of transition matrix elements. In such experiments, lattice depths are often measured by exposing an ultracold atomic gas to a series of off-resonant laser-standing-wave pulses, and fitting theoretical predictions for the fraction of atoms found in each of the allowed momentum states by time of flight measurement, after some number of pulses. We present a full analytic model for the time evolution of the atomic populations of the lowest momentum-states, which is sufficient for a ``weak''� lattice, as well as numerical simulations incorporating higher momentum states for both relatively strong and weak lattices. Finally, we consider the situation where the initial gas is explicitly assumed to be at a finite temperature.
\end{abstract}

\maketitle

\section{Introduction}
Precision measurement of optical lattice \cite{bec_in_optical_lattice_morsch_2006} depths is important for a broad range of fields in atomic and molecular physics \cite{danzl_molecules_in_lattice_2009,kotochigova_controlling_molecules_lattice_2006}, most notably in atom interferometry \cite{Cronin_2009_atom_interferometry,atom_interferometry_2006}, many body quantum physics \cite{bloch_many_body_physics_lattices_2008,gyuboong_kagome_lattice_2012}, accurate determination of transition matrix elements \cite{mitroy_safronova_matrix_elements_2010,arora_tuneout_wavelengths_2011,henson_tuneout_helium_2015,safronova_rb_matrix_elements_2015,clark_bec_matrix_elements_2015}, and, by extension, ultraprecise atomic clocks \cite{safronova_bbr_2011,sherman_polarizability_lattice_clock_2012}. Lattice depth measurement schemes include methods based on parametric heating \cite{friebel_parametric_heating_1998}, Rabi oscillations \cite{ovchinnikov_rabi_oscillations_1999}, and sudden lattice phase shifts \cite{cabreragutierez_sudden_phase_shift_2018}. The most commonly used scheme is Kapitza--Dirac scattering \cite{cahn_kaptizadirac_1997}, where an ultracold atomic gas is exposed to a pulsed laser standing wave and theoretical predictions for the fraction of atoms found in each of the allowed momentum states are fitted to time of flight measurements 
\cite{birkl_bragg_scattering_optical_lattice_1995,gyuboong_kagome_lattice_2012,cheiney_matterwave_scattering_2013,gadway_kapitzadirac_2009}. However, when determining the matrix elements of weak atomic transitions, the lattice depths involved are correspondingly small ($V\protect{\lesssim}0.01E_\mathrm{R}$ for any atom, here $V$ is the lattice depth and $E_\mathrm{R}$ is the laser recoil energy), such that signal-to-noise considerations become an issue \cite{schmidt_rb87_tuneout_2016}.

Recently, the work of Herold \textit{\textit{et al}}.\ \cite{herold_matrix_elements} and Kao \textit{\textit{et al}}.\ \cite{kao_tuneout_dysprosium} has suggested that this complication can be mitigated by using multiple laser standing wave pulses, alternating each with a free evolution, such that each alternating stage has a duration equal to half the Talbot time \cite{deng_matterwave_dispersion_talbot_1999,Kanem_2007_higher_order_quantum_resonances,Ryu_2006_higher_order_quantum_resonances_BEC}. With each pulse, population in the first diffraction order is coherently increased, improving contrast relative to the zeroth order.\footnote{In practice, this additive effect is only maintained for a certain number of pulses set by the lattice depth, as we discuss in section \ref{higherdiffractionorders}.}

The modeling approach taken in \cite{herold_matrix_elements,kao_tuneout_dysprosium} is valid for a weak lattice which is pulsed a small number of times, corresponding to the ``weakly-diffracting limit''. Following description of our model system and its general time evolution in section \ref{system}, in section \ref{twostateanalytics} we present a full analytic model for the time evolution of the atomic populations of the zeroth and first diffraction orders; this is sufficient for a ``weak'' lattice. In section  \ref{higherdiffractionorders} we present numerical simulations incorporating higher momentum states at both large and small lattice depths $V$ (``small'' is taken to mean when $V$ is less than a tenth of the recoil energy $E_\mathrm{R}$), which we compare for typical experimental values. We also explore the role of finite-temperature effects in such experiments (section \ref{finitetemperature}), and present our conclusions in section \ref{conclusion}.

\section{Model system: BEC in an optical lattice}
\label{system}
\subsection{Alternating Hamiltonian evolutions}
\begin{figure}[t]
\includegraphics{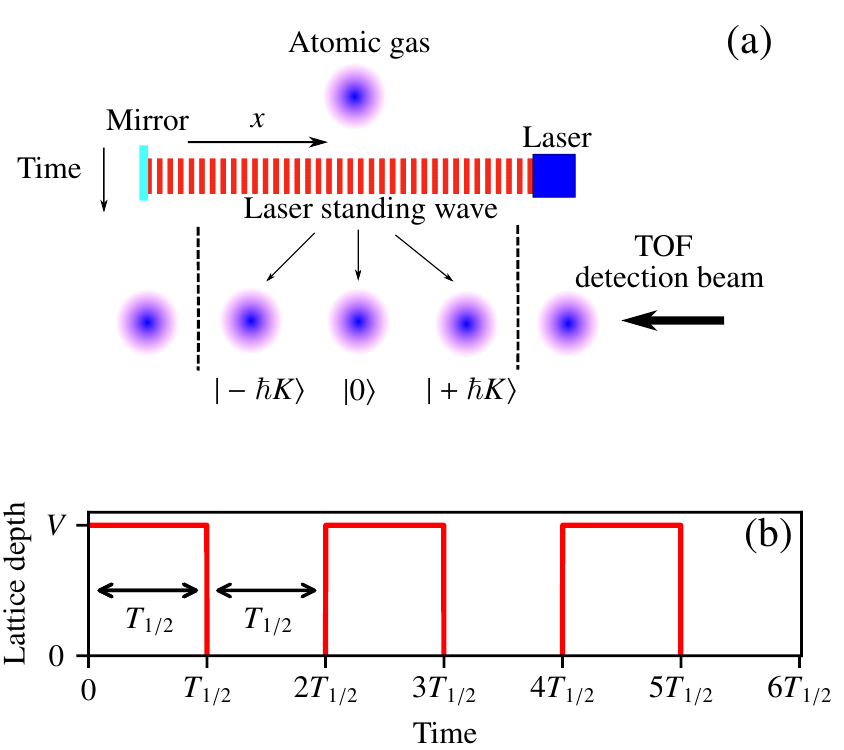}
\caption{\label{fig:sysdiagram}(Color online) Diagram of a multi-pulse atom-diffraction setup. (a) shows a cold atomic gas subjected to multiple lattice pulse evolution sequences, before a time of flight beam measures the atomic population in each of the allowed momentum states, (b) shows the modulation of the lattice depth in time, where $V$ is the lattice depth (dimensions of energy) when the standing wave pulse is on, and $T_{1/2}$ is the Talbot time as defined in Eq.\ (\ref{halftalbottime}). For simplicity, the laser standing wave has been oriented orthogonally to the gravitational direction, however we note that this is equivalent to a vertically oriented system in which a phase-shifter element is used to introduce a time dependent phase on the standing wave, which is tuned to cancel out gravity \cite{godun_2000_accelerator_modes,beswick_et_al}.
}
\end{figure}

We consider an atomic Bose--Einstein condensate (BEC) with interatomic interactions neglected.\footnote{The quantum degeneracy is not important in our analysis, as the requirement is simply for a very narrow initial momentum spread.} This can be achieved experimentally by exploiting an appropriate Feshbach resonance \cite{inouye_andrews_1998,kohler_goral_2006,gustavsson_haller_2008,molony_gregory_2014}, or by allowing the cloud to expand adiabatically \cite{jamison_atomic_interactions_interferometry_bec_2011}. Working in this regime means that we need only consider the single-particle dynamics of each atom. The optical lattice laser is far off resonance such that we consider the atomic center of mass motion only \cite{meystre_atom_optics_2001}, and we consider the atoms to be periodically perturbed by a 1d optical lattice, alternated with a free evolution  \cite{beswick_et_al}. The atomic center of mass dynamics are then alternatingly governed by the following Hamiltonians:
\begin{subequations}
\begin{align}
\hat{H}_{\mathrm{Latt}}& = \frac{\hat{p}^2}{2M} - V \cos(K\hat{x}), 
\label{Hlatt}\\
\hat{H}_{\mathrm{Free}}& = \frac{\hat{p}^2}{2M}, 
\label{Hfree}
\end{align}
\end{subequations}
where $\hat{p}$ is the 1d momentum operator in the $x$ direction (see Fig \ref{fig:sysdiagram}), $\hat{x}$ is the associated position operator, $M$ is the atomic mass, and $V$ the lattice depth\footnote{It is conventional to define the lattice depth with respect to a potential of the form $U_0\sin^2(Kx/2)$. In this work we refer to the lattice depth as $V=-U_0/2=-\hbar\Omega^2/8\Delta$ for a laser Rabi frequency $\Omega$ and detuning $\Delta\equiv \omega_L - \omega_0$.} (dimensions of energy) of a lattice with wavenumber $K$ ($K=2K_\mathrm{L}$, where $K_\mathrm{L}$ is the laser wavenumber) \cite{saunders_halkyard_gardiner_challis_2009, zheng}. 

As stated in the introduction, Herold \textit{et al}.\ \cite{herold_matrix_elements} and Kao \textit{et al}.\ \cite{kao_tuneout_dysprosium} proposed that when measuring very small lattice depths ($V\protect{\sim}0.01E_\mathrm{R}$, here $E_\mathrm{R}=\hbar^2 K^2/8M$), the signal can be optimized by both the lattice pulse and free evolution having a duration equal to the half Talbot time \cite{arnold_carson_talbot_time},
\begin{equation}
T_{1/2} = \frac{2 \pi M}{\hbar K^2}.
\label{halftalbottime}
\end{equation}
This is half the full Talbot time, which is the elapsed time for which the free evolution operator [generated by Eq.\ (\ref{Hfree})] collapses to the identity when applied to a momentum state that is an integer multiple of $\hbar K$.\footnote{For an initially zero-temperature gas, these conditions yield an \textit{antiresonance\/} in the quantum $\delta$-kicked particle (the momentum width of the gas is bounded, and alternates in time between two values) 
\cite{phase_noise_dkr_white_ruddell_hoogerland,Kanem_2007_higher_order_quantum_resonances,Ryu_2006_higher_order_quantum_resonances_BEC,Szriftgiser_2002_sub_fourier_resonances,Williams_2004_diffusion_resonances,Duffy_2004_early_time_diffusion_BEC,BEC_Ullah,saunders_halkyard_challis_gardiner_2007,saunders_halkyard_gardiner_challis_2009,power_law_behaviour_Halkyard,Oskay_2000}. }
\subsection{Time evolution}
The time-periodicity of the system admits a Floquet treatment \cite{saunders_halkyard_challis_gardiner_2007}; the time evolution of an initial state $| \psi(t=0) \rangle$ for $N$ successive lattice-pulse sequences is given by repeated applications of the system Floquet operator $\hat{F}$ to the initial state, i.e., $|\psi(t=N) \rangle = \hat{F}^N |\psi (t= 0)\rangle$. 

We determine the relevant $\hat{F}$, governing a lattice pulse of duration $T_{1/2}$ [Eq.\ (\ref{halftalbottime})], followed by a free evolution of the same duration, straightforwardly from the time evolution operators generated by Eqs.\ (\ref{Hlatt}) and (\ref{Hfree}). The spatial periodicity of the laser standing wave  also enables us to invoke Bloch theory \cite{ashcroft_mermin}. Recasting the momentum operator $\hat{p}$ such that:
\begin{subequations}
\begin{align}
(\hbar K)^{-1}\hat{p}  & = \hat{k} + \hat{\beta},
\label{Eq:MomentumParam}
\\
\hat{k}|(\hbar K)^{-1}p = k + \beta\rangle 
& = k|(\hbar K)^{-1}p = k + \beta\rangle,
\label{keigenstates}
\\
\hat{\beta}|(\hbar K)^{-1}p = k + \beta\rangle 
& = \beta|(\hbar K)^{-1}p = k + \beta\rangle,
\label{kbetaeigenstates}
\end{align}
\label{momentumeigenstatesdefinition}
\end{subequations}
with $k \in \mathbb{Z}$ and $\beta \in [-1/2,1/2)$ \cite{bach_burnett_d'arcy_gardiner_2005}, we elucidate that the total dimensionless momentum $(\hbar K)^{-1}p$ associated with a single plane wave is the sum of $k$, the discrete part, and $\beta$ as the continuous part or \textit{quasimomentum}, which is a conserved quantity. Hence, only momentum states separated by integer multiples of $\hbar K$ are coupled \cite{kicked_rotor_wigner,beswick_et_al}. Within a single quasimomentum subspace, the system Floquet operator can therefore be written:
\begin{equation}
\begin{split}
\hat{F}(\beta) = & \hat{F}(\beta)_{\mathrm{Free}}\hat{F}(\beta)_{\mathrm{Latt}} = \exp
\left(
-i\left[
\frac{\hat{k}^{2}+2\hat{k}\beta}{2}
\right]2\pi
\right)
\\& \times
\exp
\left(
-i\left[
\frac{\hat{k}^{2}+2\hat{k}\beta}{2}-V_{\mathrm{eff}}\cos(\hat{\theta})\right]2\pi
\right),
\label{floquetfinitedurationk}
\end{split}
\end{equation}
where $V_{\mathrm{eff}}= VM/\hbar^2 K^2$ is the dimensionless lattice depth, $\hat{\theta}=K \hat{x}$ and the rescaled half Talbot time is equal to $2\pi$.\footnote{In generality Eq.\ (\ref{floquetfinitedurationk}) should include the operator $\hat{\beta}$, however, restricting our analysis to states within a single quasimomentum subspace, $\beta$ is a scalar value, and relative phases depending solely on $\beta$ can be neglected.}
Using Eq.\ (\ref{floquetfinitedurationk}) to calculate $|\psi(t=N) \rangle=\sum_j c_j(N) | k = j \rangle$, the population in each discrete momentum state $| k = j \rangle$ after $N$ pulses is given by the absolute square of the individual coefficients $P_j(N)=|c_j(N)|^2$. In this paper we employ both the well-known split-step Fourier approach \cite{daszuta_andersen_2012,beswick_et_al}, and matrix diagonalization in a truncated basis \cite{herold_matrix_elements, standing_wave_numerical_diagonalization} to determine $|\psi(t=N) \rangle$ beyond the weakly-diffracting limit, as well as an analytic approach in the weakly-diffracting case.

\section{Analytic results in a two-state basis}
\label{twostateanalytics}
For an initially zero-temperature gas ($\beta=0$) subjected to a small number of pulses from a shallow lattice, a useful approximation is to assume that no population is diffracted into momentum states with $|p| > \hbar K$, the so-called ``weakly-diffracting limit''. Mathematically, this regime corresponds to the time evolution of an initial state $|\psi(t=0) \rangle = | k=0 \rangle$ in a space spanned only by the $| k=-1 \rangle$, $| k=0 \rangle$ and $| k=1 \rangle$ states of the $\beta=0$ quasimomentum subspace. 

The symmetry of the lattice and free evolution Hamiltonians about $| k=0 \rangle$ guarantees that, for our chosen initial state, the population diffracted into the $| k=1 \rangle$ state is identical to that diffracted into the $| k=-1 \rangle$ state. We therefore express the system Hamiltonians (\ref{Hlatt}), (\ref{Hfree}) as matrices in the truncated momentum basis:
\begin{subequations}
\label{trunc_basis}
\begin{align}
|0\rangle &= |k=0\rangle = \left( \begin{array}{c}
0\\
1\\
0
\end{array}
\right), \label{column0} \\
|+\rangle&=\frac{1}{\sqrt{2}}(|k=1\rangle + |k=-1\rangle) = \left( \begin{array}{c}
1\\
0\\
0
\end{array}
\right), \label{columnminus} \\
|-\rangle&=\frac{1}{\sqrt{2}}(|k=1\rangle - |k=-1\rangle) = \left( \begin{array}{c}
0\\
0\\
1
\end{array}
\right)\label{columnplus},
\end{align}
\end{subequations}
yielding the following $3\times3$ matrix representation of the lattice Hamiltonian:
\begin{equation}
H^{3\times3}_{\mathrm{Latt}}=
\begin{pmatrix}
    1/2   & -V_\mathrm{eff}/\sqrt{2} & 0  \\
     -V_\mathrm{eff}/\sqrt{2}  & 0 & 0  \\
     0   & 0 & 1/2 
\end{pmatrix}.
\label{decoupled}
\end{equation}

There is no coupling between the $|0\rangle$ state and the antisymmetric $|-\rangle$ state. Hence, for an initially zero-temperature gas, there is no population transfer into the $|-\rangle$ state for all time. The relevant basis is therefore two-dimensional, with basis states $|0\rangle_2\equiv\tvect{0}{1}$ and $|+\rangle_2\equiv\tvect{1}{0}$. We use these to represent Eq.\ (\ref{Hlatt}) as the $2\times2$ matrix: 
\begin{equation}
H^{2\times2}_{\mathrm{Latt}}=
\begin{pmatrix}
    1/2  &  -V_\mathrm{eff}/\sqrt{2} \\
      -V_\mathrm{eff}/\sqrt{2}  &  0 
\end{pmatrix}.
\label{rabi}
\end{equation}
We recognize Eq.\ (\ref{rabi}) as a Rabi matrix, the eigenvalues and normalized eigenvectors of which are well known \cite{barnett_radmore_methods_qo_1997}. We use these to calculate the populations after $N$ pulses of the $|0\rangle$ and $|+\rangle$ states [$P_0(N,V_\mathrm{eff})$ and $P_+(N,V_\mathrm{eff})$, respectively]:
\begin{subequations}
\begin{align}
P&_0(N,V_\mathrm{eff}) =1- A \sin^2(N\phi/2), \label{pzero}\\
P&_+(N,V_\mathrm{eff}) = A \sin^2(N\phi/2), \label{pplus} \\
A &= \frac{8 V_\mathrm{eff}^2 \sin^2\left(\pi \sqrt{1+8V_\mathrm{eff}^2}/2\right)}{8 V_\mathrm{eff}^2 + \cos^2\left(\pi \sqrt{1+8V_\mathrm{eff}^2}/2\right)} \label{amplitudea},\\
\phi &= 2 \, \arctan\left( \frac{\sqrt{8V_\mathrm{eff}^2 + \cos^2\left(\pi \sqrt{1+8V_\mathrm{eff}^2}/2\right)}}{\sin\left(\pi \sqrt{1+8V_\mathrm{eff}^2}/2\right)}\right)\label{frequencyphi},
\end{align}
\end{subequations}
as explicitly derived in Appendix \ref{analyticsappendix}.

\begin{figure}[t]
{\centering
\includegraphics{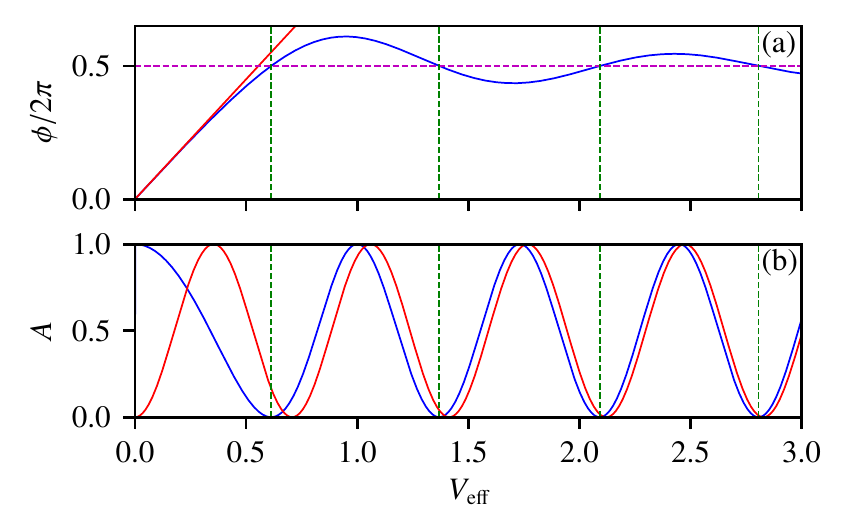}
\caption{(Color online) Plot of the variation of $\phi/2\pi$, (a), and the amplitude A, (b), versus $V_\mathrm{eff}$, all quantities are dimensionless. The blue curves [beginning at $\phi/2\pi=0$ for (a), and $A=1$ for (b)] give the full analytic form for each expression, corresponding to Eqs.\ (\ref{frequencyphi}) and (\ref{amplitudea}) respectively. The solid red lines show our linear approximation to $\phi$ for $V_\mathrm{eff} \ll 1$, $\phi \approx 4 \sqrt{2} V_\mathrm{eff}$ [the straight line of (a)], and our limiting value of $A$ for $V_\mathrm{eff} \rightarrow \infty$, $A=\sin^2(\sqrt{2} \pi V_\mathrm{eff})$ [the lowermost curve of (b)]. The horizontal dashed line in (a) appears at $\phi=\pi$, which is a physically relevant value about which $\phi$ oscillates beyond its first turning point. The vertical lines correspond to the points where $\phi=\pi$, and $A=0$, both of which always occur simultaneously.\label{amplitudephifig} 
}
}
\end{figure}

From Eqs.\ (\ref{pzero}) and (\ref{pplus}), we see that in the weakly-diffracting limit $P_0$ and $P_+$ oscillate sinusoidally with the number of pulses $N$, and are entirely characterized by an amplitude $A$ and a ``frequency'' $\phi$, both of which depend solely on the dimensionless lattice depth $V_\mathrm{eff}$. We note the similarity to the result reported in \cite{gadway_kapitzadirac_2009} for single pulse diffraction. We display the variation of $A$ and of $\phi$ versus $V_\mathrm{eff}$ in Fig.\ \ref{amplitudephifig}\footnote{Note that when explicitly evaluating Eq.\ (\ref{frequencyphi}), it is desirable to use the ``Atan2'' numerical routine in, e.g., Python. This ensures that the sign of the argument is taken into account, which avoids singularities in the frequency.}; 
$\phi$ initially increases approximately linearly with $V_\mathrm{eff}$, meaning that over a sufficiently small range of lattice depths, we should expect to see an approximate universality in the population dynamics when the time axis is scaled by $V_\mathrm{eff}$ (we explore this scaling in Section \ref{higherdiffractionorders}). In the limit where $V_\mathrm{eff} \rightarrow 0$, it follows that $\phi=4\sqrt{2}V_\mathrm{eff}$ (see Appendix \ref{limitsappendix}), depicted by the solid straight line plotted in Fig. \ref{amplitudephifig}(a). Substituting this result into Eq.\ (\ref{pplus}) and expanding the corresponding Taylor series to leading order, we recover the familiar quadratic dependence of Herold et al.\ \cite{herold_matrix_elements,kao_tuneout_dysprosium} (see Appendix \ref{heroldlimit}): 
\begin{equation}
P_+=8 N^2 V_\mathrm{eff}^2\propto N^2.
\label{herold_quadratic}
\end{equation}
The validity of this result is subject to $N\phi(V_\mathrm{eff})/2\ll1$. Increasing $V_\mathrm{eff}$ beyond this regime, $A$, which decreases steadily in the range of linearity of $\phi$, first reaches a node at $V_\mathrm{eff}=\sqrt{3}/(2\sqrt{2})\simeq0.612$, and afterwards at all points where $V_\mathrm{eff} = \sqrt{4m^2-1}/(2\sqrt{2})$, $m \in \mathbb{Z}^+$, depicted by the vertical dashed lines of Fig \ref{amplitudephifig}. Physically, these values of $V_\mathrm{eff}$ correspond to there being no pulse-to-pulse population transfer out of the $|k=0\rangle$ state, at least in the weakly-diffracting limit. As shown in Appendix \ref{limitsappendix}, $\phi=\pi$ at those values of $V_\mathrm{eff}$ where $A$ has a node, visualised by the intersection of the vertical and horizontal dashed lines in Fig \ref{amplitudephifig}(a).

In the limit where $V_\mathrm{eff} \rightarrow \infty$, $\phi=\pi$ whenever $V_\mathrm{eff}=n/\sqrt{2}$, with an overall oscillatory behavior of ever-decreasing amplitude around this value, while $A$ takes on the form of a sinusoidal oscillation: $A=\sin^2(\sqrt{2}\pi V_\mathrm{eff})$.
\section{Incorporating higher diffraction orders}
\label{higherdiffractionorders}
\subsection{Numerical simulations for a large momentum basis}
Having obtained analytic results for the time-evolved populations in the weakly-diffracting limit, we test their domain of validity by using standard numerical techniques to compute the full momentum distribution of the system, and sampling the population in the $|k=0\rangle$ state, $P_0$. We follow the same approach as \cite{saunders_halkyard_challis_gardiner_2007,daszuta_andersen_2012} and work within the momentum basis. The action of the Floquet operator (\ref{floquetfinitedurationk}) on the total state of the system, $|\psi\rangle$, is calculated by a split-step Fourier method, on a basis of $2048$ momentum states, which is exhaustive for any practical purpose.

\begin{figure}[t]
{\centering
\includegraphics{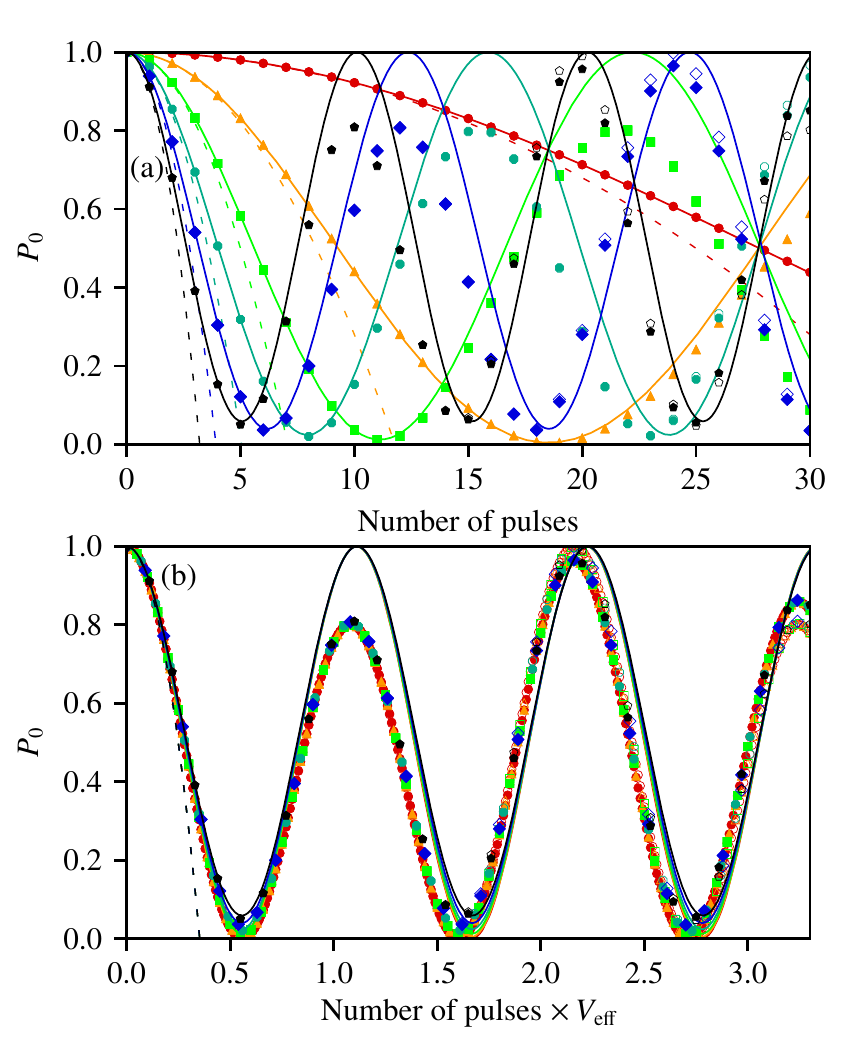}
\caption{(Color online) (a): Plot of population in the $|k=0\rangle$ state, $P_0$, versus number of pulses, as calculated in a truncated momentum basis with $|k|\leq3$ by numerical diagonalization (hollow markers), and a 2048 state basis using a split-step Fourier method (solid markers). The solid lines correspond to the analytic solution for $P_0$ in a two state basis, as given by Eq.\ (\ref{pzero}), while the dashed lines represent the quadratic solution of Herold et al.\ [Eq.\ (\ref{herold_quadratic})]. Each set of markers corresponds to a fixed value of the effective lattice depth ranging from the slowest-oscillating curve at $V_\mathrm{eff}=0.01$ to the fastest oscillating one at $V_\mathrm{eff}=0.11$ in steps of $0.02$. (b): Reproduction of (a), with the number of pulses axis scaled by the dimensionless lattice depth $V_\mathrm{eff}$ to reveal an approximate universal curve both in the analytics and the numerical simulations. The data have been extended to span the full range of the horizontal axis. The universal curve reveals a drop in the amplitude of $P_0$ as calculated by the full numerics at the first revival, which is not reproduced by the analytics, but is reproduced in the truncated momentum basis. In (b), the oscillation frequency of the numerical curve increases compared to that of the analytic result as the number of pulses or the lattice depth is increased. After three half-oscillations on the universal curve, the truncated basis result begins to deviate appreciably from the full numerics.\label{fig:universalcurve} 
}
}
\end{figure}

In Fig.\ \ref{fig:universalcurve} we compare the analytic results of Eqs.\ (\ref{pzero},\ref{pplus},\ref{amplitudea},\ref{frequencyphi}) to this exact numerical calculation  for fixed values of the effective lattice depth $V_\mathrm{eff}$. From Fig.\ \ref{fig:universalcurve}(a) we see that the sinusoidal character of the analytic result for $P_0$ is revealed for higher values of $V_{\mathrm{eff}}$, as well as a similar oscillatory behavior in the exact numerics. Naively, we may say that increasing $V_\mathrm{eff}$ gives rise to a greater deviation of the exact numerics from the analytics. This is true when comparing over a fixed number of pulses, however we can use our argument that there is an approximate universality in $V_\mathrm{eff}$ and the number of pulses (see section \ref{twostateanalytics}) to clarify this statement by means of the universal curve displayed in Fig.\ \ref{fig:universalcurve}(b). This clearly shows that the universality holds approximately for the exact numerics also, and that the analytics cease to agree with the exact numerics at approximately the same point on the universal curve, regardless of the value of $V_\mathrm{eff}$ in the chosen range. Hence, more completely, the analytics are sufficient to understand the system provided the product of the number of pulses and effective lattice depth is sufficiently small. We note specifically that there is a frequency drift which increases along the curve, and a marked reduction in amplitude of the exact numerics as compared to the analytics at its first revival. Both features appear due to leakage of population into momentum states with $|p|>\hbar K$, and inform our discussion of the range of validity of the weakly-diffracting limit taken in previous work. Indeed, the quadratic result of Herold et al. [Eq.\ (\ref{herold_quadratic}), shown as dashed lines in Fig.\ \ref{fig:universalcurve}] deviates from the exact numerics at a significantly smaller value of $NV_\mathrm{eff}$ than our exact analytic result for two diffraction orders.

\begin{figure*}[t]
{\centering
\includegraphics{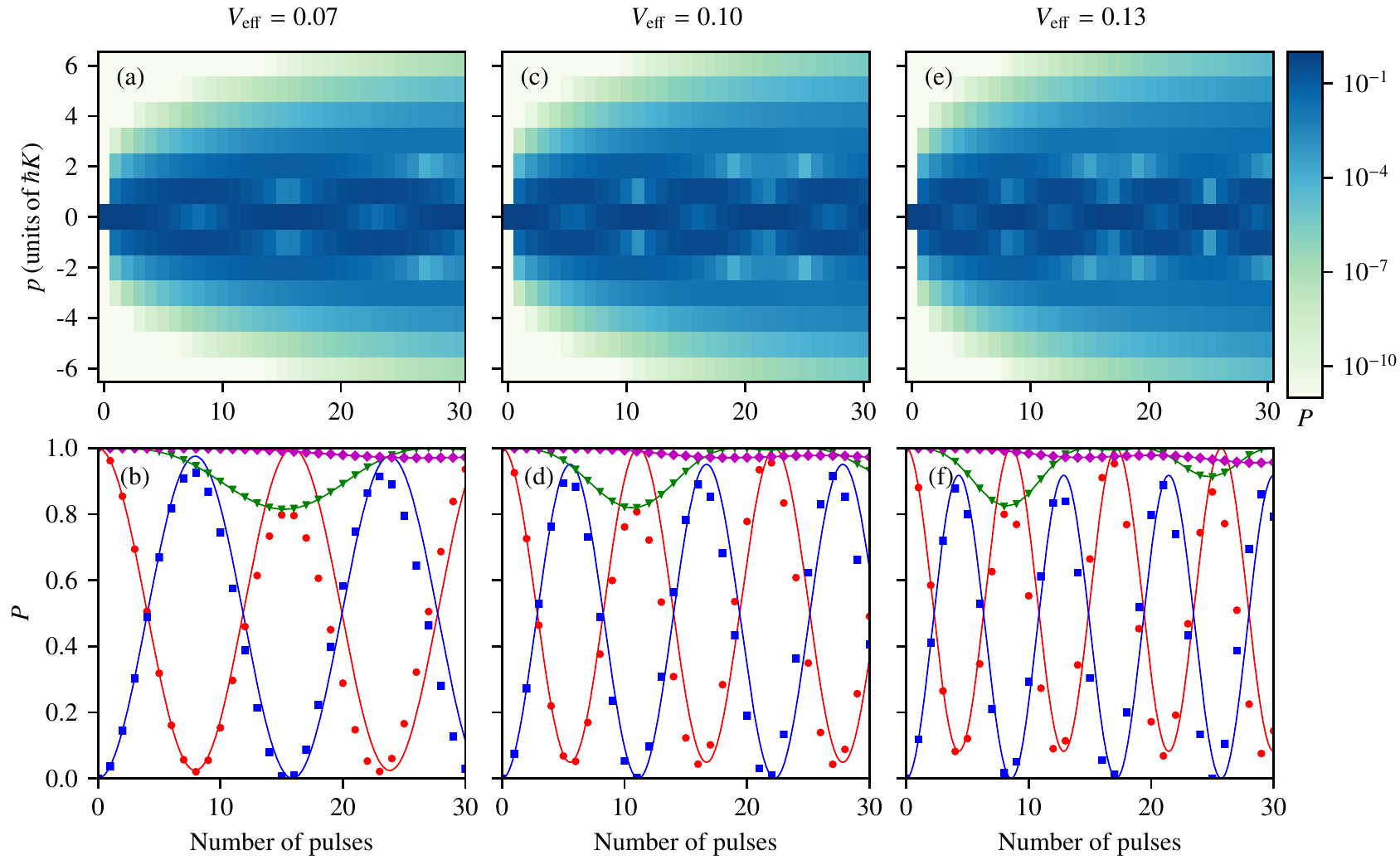}
\caption{(Color online) Comparison between population dynamics for differing values of the dimensionless lattice depth $V_\mathrm{eff}$, as computed by exact numerics and the two-state analytic model of Eqs.\ (\ref{pzero}) and (\ref{pplus}). Row 1 [(a), (c), (e)] comprises momentum distributions versus the number of lattice pulses for an initially zero-temperature gas in a basis of 2048 momentum states. Each false-color plot shows the time evolved population in the first 13 momentum states ($|k| < 6$), to be read on the colorbar to the right.  A cutoff population value of $P_\mathrm{cutoff}=10^{-11}$ has been applied to each population distribution to accommodate the log scale. This illustrates that for this choice of parameters, the amount of population diffracted into momentum states with $|p| > 3\hbar K$ is negligible. Row 2 [(b), (d), (f)] shows firstly, slices through the momentum distribution corresponding to the population in the $k=0$ state, $P_0$, (red circles) and the $|p|=\hbar K$ states, $P_{\pm1}$, (blue squares), to which our two-state analytic model is compared (red and blue solid lines respectively). To clarify the drop in amplitude in the first revival of $P_0$, the green triangles have been added, which correspond to $1-P_{\pm2}$ and almost intersect the red circles corresponding to $P_0$, indicating that the overwhelming majority of the population which has left $P_0$ at this point, has in fact been diffracted into the $|k|=2$ states. At the second revival, the two sets of points are further apart. Population leakage into the $|p|=3\hbar K$ states, corresponding to the magenta diamonds, which represent $1-P_{\pm3}$, explains this effect. Solid lines have been added as a guide to the eye. Each column corresponds to a fixed value of $V_\mathrm{eff}$, [(a),(b)] $V_\mathrm{eff}=0.07$, [(c),(d)] $V_\mathrm{eff}=0.10$, [(c),(d)] $V_\mathrm{eff}=0.13$.
\label{fig:highermomentumstates} 
}
}
\end{figure*}

In \cite{herold_matrix_elements,kao_tuneout_dysprosium}, the regime in which the weakly-diffracting limit is satisfied (recast in our system of variables) is given by $NV_\mathrm{eff}\ll1/4$. Though this inequality places an upper bound on the allowed value of $NV_\mathrm{eff}$, it is reasonable to ask at what point is $NV_\mathrm{eff}$ ``much smaller'' than $1/4$? By inspection of Fig.\ \ref{fig:universalcurve}(b), we can see that at $NV_\mathrm{eff}=1/4$, there is still excellent agreement between our analytics and exact numerics. We calculate the RMS difference between our analytics and full numerics \cite{hughes_hase_measurements_uncertanties_2010} at this point over the range of chosen lattice depths (defined as RMS=$[\sum_{j=1}^{\mathcal{N}} \{P_0(N,V_\mathrm{eff})_j-P_{0(\mathrm{Numerical})}(N,V_\mathrm{eff})_j \}/\mathcal{N}]^{1/2}$, where $\mathcal{N}$ is the number of lattice depth values) to be $0.0011$ (deviation at the 0.1\% level). The corresponding quadratic result deviates at the 42\% level.\footnote{In practice, the discretization of the time axis in the number of pulses means that we cannot generally assume that any data points from the full numerics will fall at the exact value $NV_\mathrm{eff}=1/4$, and so we have chosen the data closest to this point in our calculation of the RMS.} The point at which leakage into higher momentum-states first becomes appreciable is $NV_\mathrm{eff}\sim1/2$, with an RMS of $0.0043$ (deviation at the 0.4\% level). Though this is clearly sufficiently small to still be considered within the range of validity of the weakly-diffracting limit, beyond $NV_\mathrm{eff}\sim1/2$, where the RMS becomes larger, we must incorporate higher momentum-states. This motivates the question of how many momentum states are necessary to include for such a model to be useful for a reasonable choice of experimental parameters.

Figures \ref{fig:highermomentumstates} (a,c,e) show a selection of momentum distributions for a range of values of $V_\mathrm{eff}$ as calculated by the full numerics, showing momentum states up to $|p|\leq6\hbar K$, with Figs.\ \ref{fig:highermomentumstates} (b,d,f) showing corresponding slices through the momentum distributions. The log scale makes clear that there is very little population leakage into momentum states with $|p|>3\hbar K$ for the chosen values. Instead we see that there are pronounced oscillations in population between the $|p|=0$ and $|p|=\hbar K$ states, which are modulated by population leakage into the $|p|=2\hbar K$ states, and to a lesser extent the $|p|=3\hbar K$ states. By inspection of the lattice Hamiltonian in the momentum basis, this can be explained by the decrease in magnitude of the off-diagonal coupling terms with state number. In fact, the decrease in amplitude at the first revival in Fig.\ \ref{fig:universalcurve}(b) is almost entirely due to population leakage into the $|p|=2\hbar K$ states, suggesting that a model incorporating only $n=5$ momentum states should be sufficient to capture the dynamics, up to at least $V_\mathrm{eff}=1.1$. 

\subsection{Small momentum bases of dimension $\boldsymbol{>2}$}
To incorporate higher momentum-states we numerically diagonalize Eqs.\ (\ref{Hlatt}) and (\ref{Hfree}), in a truncated basis of $n$ momentum states, and propagate the time-evolution using the procedure described in Appendix \ref{numerical_diagonalization}. Our analysis in the previous section suggests that simulations using a basis of $n=5$ momentum states ought to be sufficient for practical purposes. Corresponding results are shown by the hollow markers in Fig.\ \ref{fig:universalcurve}(b). The five state model is an order of magnitude more accurate than the analytics at $NV_\mathrm{eff}=1/4$ and $NV_\mathrm{eff}=1/2$, with RMS differences with respect to the full numerics of 0.00018, and 0.00011 respectively. As expected, the decrease in amplitude at the second revival on the universal curve is reproduced by this approach, but is clearly also valid over a larger range, up to the fourth turning point ($NV_\mathrm{eff}\sim1.6$, RMS deviation 0.0022), beyond which the model begins to overestimate and then underestimate the exact numerical result. 

This difference appears as a result of the basis truncation, as population leakage into states with $|p|\geq 5\hbar K$ is explicitly not possible in this model, though it should be noted that this effect would only be relevant to experiments performed using a very large effective lattice depth. An attractive feature of the five state model is that it can in principle be solved analytically for the time-evolution of the populations, which could be fit to experimental data to extract more accurate lattice depths.

\section{Finite-temperature response}
\label{finitetemperature}
\begin{figure}[t]
{\centering
\includegraphics{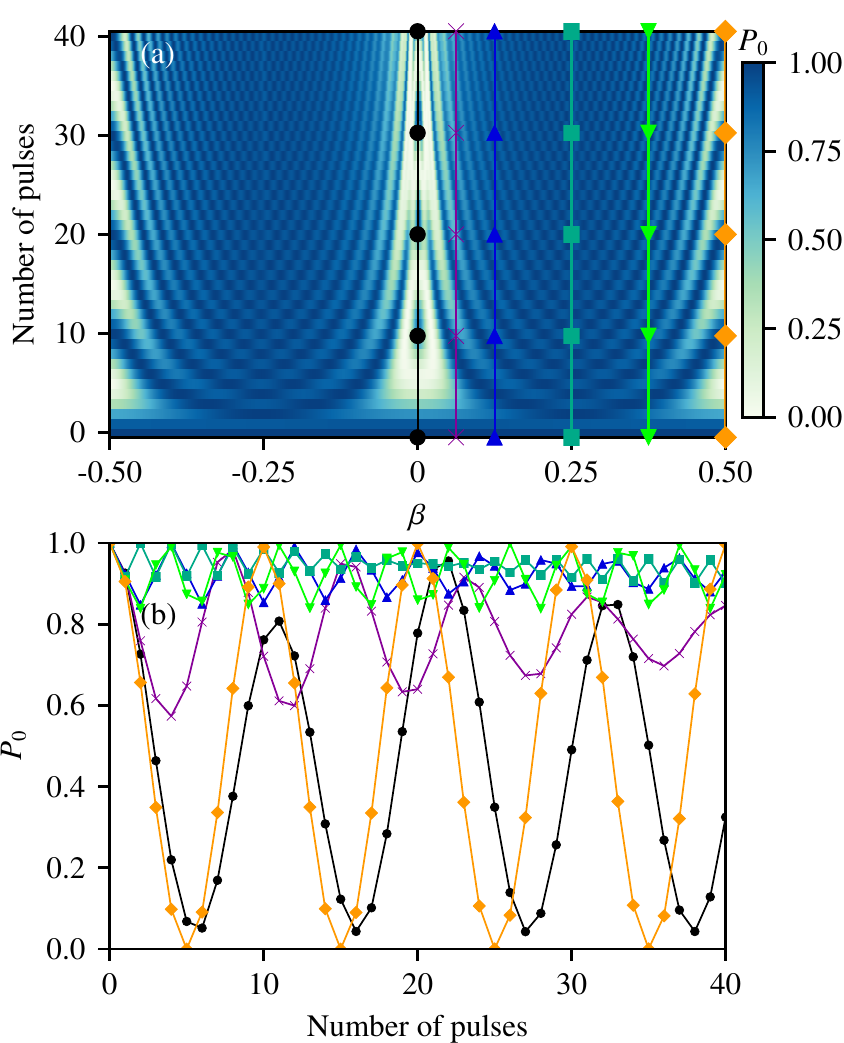}
\caption{(Color online) (a) False-color plot of the time evolution of $P_0$ as computed in a basis of 2048 momentum states for values of the dimensionless quasimomentum $\beta$ [see Eq. (\ref{momentumeigenstatesdefinition})] ranging from $\beta = -0.5$ to $\beta = 0.5$ in steps of $\beta =0.00025$ (4001 quasimomentum values). We have chosen a relatively large lattice depth of $V_\mathrm{eff}=0.1$ such that the different dynamical behaviors are made clear for the chosen number of pulses $N=40$\label{betacarpet}. (b) Slices taken through the quasimomentum distribution parallel to the time axis for $\beta=0,0.0625,0.125$, then increasing in increments of $\beta=0.125$ up to a maximum of $\beta=0.5$, enclosing the full range of dynamics in the $k=0$ subspace. Each vertical set of markers in (a) corresponds to the position in the quasimomentum distribution of the slices in (b), where the solid lines have been added as a guide to the eye.\label{betaplanewaves}
}
}
\end{figure}

The results presented in the previous sections are valid for a gas which is assumed to be initially at zero temperature; in practice this regime is never fully achieved, even for a BEC. To find the response of $P_0$ versus the number of pulses for a finite-temperature gas, we calculate the time evolution of $P_0$ for an ensemble of initial momentum states $|\psi(t=0)\rangle=|(\hbar K)^{-1}p = k + \beta\rangle$ according to Eq.\ (\ref{floquetfinitedurationk}), where the initial momentum is defined in a Bloch framework with $k$ and $\beta$ as free parameters. For a sufficiently cold gas [temperature $\mathcal{T}_w\lesssim(\hbar^2K^2/64k_\mathrm{B})\mathrm{K}$]\footnote{This rule of thumb is chosen such that the initial width of the momentum distribution is at most one quarter that of the first Brillouin zone.} we need only consider initial states with $k=0$ in order to capture the essential features. In this regime we choose a fixed value of the lattice depth and scan across the full range of the quasimomentum $\beta$ as the only free parameter, to find the momentum dependence in the first Brillouin zone \cite{ashcroft_mermin} displayed in Fig.\ \ref{betaplanewaves}.

Figure \ref{betaplanewaves} clearly shows the central resonance at $\beta=0$, where our zero-temperature analysis is applicable. Increasing the quasimomentum to $|\beta|=0.0625$, we see that the oscillation in $P_0$ has an amplitude of less than 50\% of that at $\beta=0$, and a substantially different frequency. Hence, the width of the central resonance is relatively narrow compared to the full width of the Brillouin zone. For an initial momentum distribution of appreciable width we must consider the surrounding structure when calculating the population dynamics, as the zero-temperature behavior will be washed out over time, or even be unresolvable altogether if the temperature is sufficiently high.

Note that for broader initial momentum distributions the dynamics will include the secondary resonances at $|\beta|=0.5$, which have a periodicity of the form $P_0(N)=\cos^2(\pi V_\mathrm{eff}N)$, such that $P_0$ varies between 0 and 1 for all $V_\mathrm{eff}$.

Having characterized the first Brillouin zone, we calculate the full finite-temperature response of $P_0$ by performing Gaussian weighting in momentum space according to a rescaled Maxwell-Boltzmann distribution:
\begin{equation}
D_{k=0}(\beta)=\frac{1}{w \sqrt{2\pi}} \exp \left(\frac{-\beta^2}{2 w^2} \right),
\label{gaussiandist}
\end{equation}
where the dimensionful temperature is given by $\mathcal{T}_{w}=\hbar^{2} K^{2} w^{2}/M k_{\mathrm{B}}$ \cite{saunders_halkyard_challis_gardiner_2007}, and $k_{\mathrm{B}}$ is Boltzmann's constant.

\begin{figure*}[t]
{\centering
\includegraphics{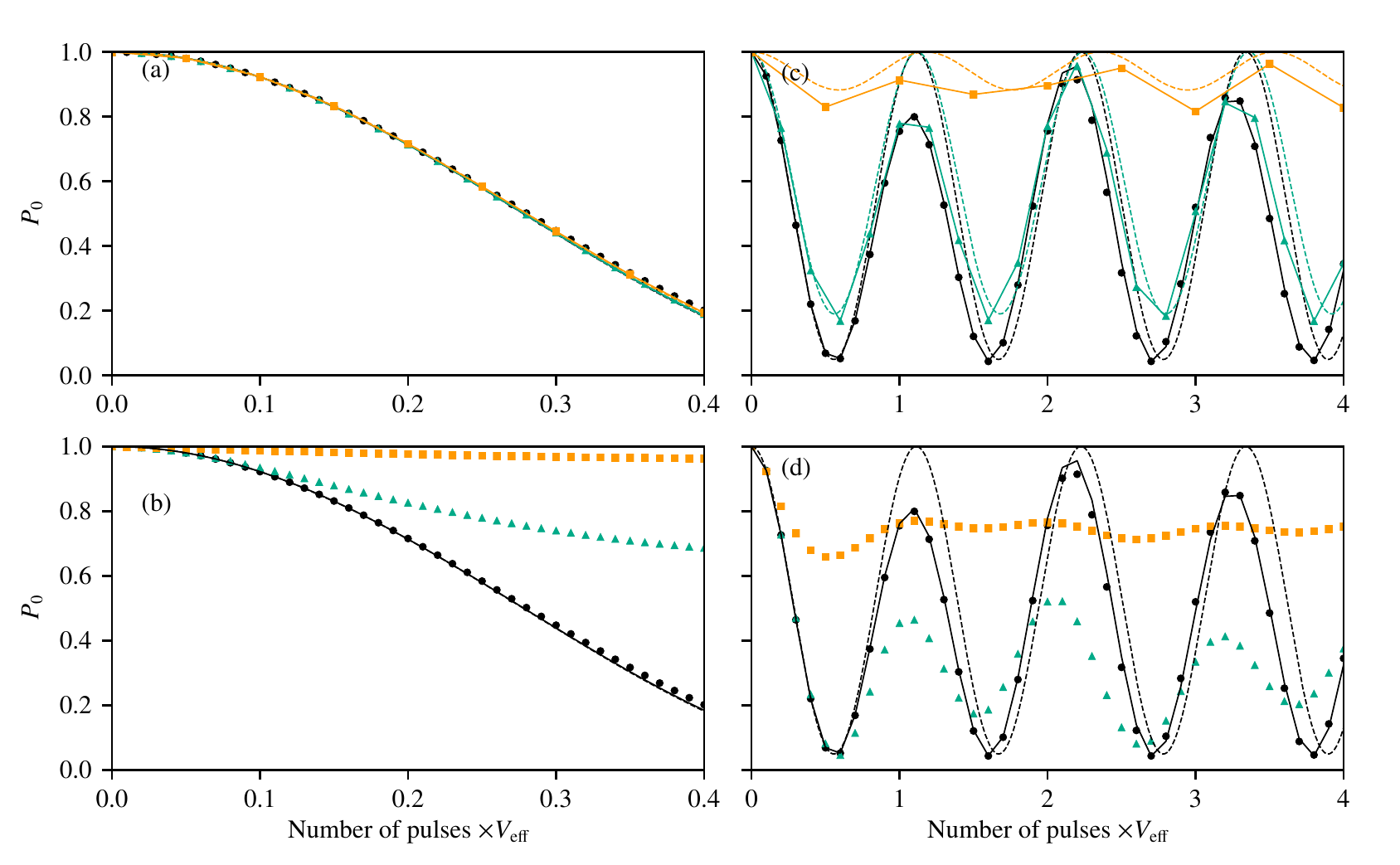}
\caption{(Color online) Plot of the finite temperature response of $P_0$ vs (number of pulses)$\times V_\mathrm{eff}$, where $V_\mathrm{eff}$ is the dimensionless lattice depth [see Eq.\ (\ref{floquetfinitedurationk})], as calculated for an ensemble of 4001 particles each evolved in a basis of 2048 momentum states. The left column [(a), (b)] corresponds to the weak-lattice regime, and the right column [(c), (d)] to the strong-lattice regime. The top row of plots [(a), (c)] shows the finite-temperature response of $P_0$ at a temperature of $w=0.00125$ for a selection of different lattice depths, $V_\mathrm{eff}=0.01,0.02,0.05$ (all curves fall on top of each other) in the weak regime (a) and $V_\mathrm{eff}=0.1,0.2,0.5$ (lower, middle and uppermost curves) in the strong regime (b). For the bottom row [(b), (d)], each set of curves and markers corresponds to the response of $P_0$ at a different temperature ($w=0.00125,0.0125,0.125$; lower, middle and uppermost curves respectively), where the effective lattice depth is kept constant at $V_\mathrm{eff}=0.1$ in the strong-lattice case and $V_\mathrm{eff}=0.01$ the weak-lattice case. In all panels, the solid lines correspond to the exact numerical result for a given lattice depth at zero temperature, while the dashed lines represent the corresponding analytic result at zero temperature in a basis of three momentum states [Eq.\ (\ref{pzero})].\label{finitetempresponse}
}
}
\end{figure*}

Figure \ref{finitetempresponse} shows the variation of $P_0$ with the number of pulses, including both the strong and weak lattice regimes, and three different values of the initial momentum distribution width $w$. In overview: in regimes where we have a weak lattice and low temperature the analytic formula is adhered to almost perfectly; in regimes where we have a weak lattice and a higher temperature we begin to see noticeable deviations, which occur for a smaller number of pulses as the temperature is increased; in regimes where we have a strong lattice and low temperature, although the analytic formula is not strongly adhered to as the lattice depth increases, the oscillation frequency appears to be reasonably robust as $V_\mathrm{eff}$ increases and the amplitude of oscillation consequently decreases; finally in the regime of strong lattice and higher temperature, the analytic formula is again only adhered to for relatively short times, with that time being dependent on the temperature.
\section{Conclusions}
\label{conclusion}
We have a zero-temperature analytic formula which yields significant insight assuming that we are working in the weakly-diffracting limit. We have shown that at zero temperature, very small basis sizes are sufficient to capture the essential features of the population dynamics outside the weakly-diffracting limit. We have explored the effects of finite temperature initial distributions, and elucidated regimes from which the lattice depths can be determined from the observed dynamics in the lowest diffraction order.

\acknowledgments
B.T.B., I.G.H., and S.A.G. thank the Leverhulme Trust research programme grant RP2013-k-009, SPOCK: Scientific
Properties of Complex Knots for support. We would also like to acknowledge helpful discussions with Charles S. Adams, Sebastian Blatt, Alexander Guttridge, Creston D. Herold and Andrew R. MacKellar.

\appendix

\section{Time evolution for 2 diffraction orders \label{analyticsappendix}}
\subsection{Floquet operator in two-state basis}
We may calculate the time evolution of the $|0\rangle$ and $|+\rangle$ state populations by first diagonalizing Eq.\ (\ref{rabi}) (reproduced here for convenience)
\begin{equation}
H^{2\times2}_{\mathrm{Latt}}=
\begin{pmatrix}
    1/2  &  -V_\mathrm{eff}/\sqrt{2} \\
      -V_\mathrm{eff}/\sqrt{2}  &  0 
\end{pmatrix},
\end{equation}
using the well known eigenvalues and normalized eigenvectors of a Rabi matrix, $E_{\pm} = (1 \pm \sqrt{1+ 8 V_\mathrm{eff}^2})/4$, and
\begin{subequations}
\begin{align}
|E_+\rangle =& \left( \begin{array}{c}
\cos(\alpha/2)\\
-\sin(\alpha/2)\end{array}
\right), \\
|E_-\rangle =& \left( \begin{array}{c}
\sin(\alpha/2)\\
\cos(\alpha/2)
\end{array}
\right),
\end{align}
\end{subequations}
respectively, where $\alpha=\arctan(2 \sqrt{2}V_\mathrm{eff})$. $H^{2\times2}_{\mathrm{Latt}}$ can then be written:
\begin{equation}
H_\mathrm{diag} = R^{\dagger}H^{2\times2}_{\mathrm{Latt}}R = 
\begin{pmatrix}
    E_+   & 0 \\
     0  & E_-
\end{pmatrix},
\end{equation}
such that $R$ is the matrix of normalized eigenvectors. This leads directly to the part of the Floquet operator governing the lattice evolution:
\begin{equation}
F_{\mathrm{Latt}}=
R^{\dagger}
\begin{pmatrix}
   e^{-2\pi i E_+ }   & 0 \\
     0  & e^{-2\pi i E_- }
\end{pmatrix}
R.
\end{equation}
Expressing $F_{\mathrm{Free}}$ in the truncated momentum basis, $|0\rangle_2\equiv\tvect{0}{1}$; $|+\rangle_2\equiv\tvect{1}{0}$, we can represent the total Floquet operator in matrix form thus:
\begin{equation}
F=F_{\mathrm{Free}}F_{\mathrm{Latt}}=
\begin{pmatrix}
   -1   & 0 \\
     0  &1
\end{pmatrix}
R^{\dagger}
\begin{pmatrix}
   e^{-2\pi i E_+ }   & 0 \\
     0  & e^{-2\pi i E_- }
\end{pmatrix}
R.
\label{floquettwostate}
\end{equation}
\subsection{Floquet evolution for a general two-level system}
Any time-evolution operator associated with a two-level system can be expressed as a $2\times2$ unitary matrix, and all unitary matrices are diagonalizable, hence we may represent such a time-evolution operator thus:
\begin{equation}
U=SU_\mathrm{diag}S^\dagger
=
\begin{pmatrix}
   v_1^+   & v_1^- \\
     v_0^+  & v_0^-
\end{pmatrix}
\begin{pmatrix}
   \lambda^+   & 0 \\
     0  & \lambda^-
\end{pmatrix}
\begin{pmatrix}
   v_1^+   & v_1^- \\
     v_0^+  & v_0^-
\end{pmatrix}^\dagger.
\end{equation}
Here $S$ is a matrix composed of the normalized eigenvectors of $U$:
\begin{equation}
\vec{v}_+ =  \left( \begin{array}{c}
v_1^+\\
v_0^+
\end{array}
\right),
\qquad
\vec{v}_- = \left( \begin{array}{c}
v_1^- \\
v_0^-
\end{array}
\right),
\end{equation}
and $\lambda^\pm$ are the corresponding eigenvalues of $U$, which have unit magnitude and so can be expressed as:
\begin{equation}
\lambda^\pm = \exp(-i \theta_\pm),
\label{lambdas}
\end{equation}
where $\theta_+$ and $\theta_-$ are phase angles to be determined. The matrix which produces $N$ successive evolutions can therefore be written:
\begin{widetext}
\begin{equation}
U= SU_\mathrm{diag}^NS^\dagger
=
\begin{pmatrix}
   v_1^+   & v_1^- \\
     v_0^+  & v_0^-
\end{pmatrix}
\begin{pmatrix}
   (\lambda^+)^N   & 0 \\
     0  & (\lambda^-)^N
\end{pmatrix}
\begin{pmatrix}
   v_1^+   & v_1^- \\
     v_0^+  & v_0^-
\end{pmatrix}^\dagger=
\begin{pmatrix}
   (\lambda^+)^N |v_1^+ |^2 + (\lambda^-)^N |v_1^- |^2  & (\lambda^+)^N v_1^+ (v_0^+)^* + (\lambda^-)^N v_1^- (v_0^-)^*  \\
   (\lambda^+)^N v_0^+ (v_1^+)^* + (\lambda^-)^N v_0^- (v_1^-)^*  & (\lambda^+)^N |v_0^+ |^2 + (\lambda^-)^N |v_0^- |^2
\end{pmatrix}.
\label{Uexpanded}
\end{equation}
\end{widetext}
Suppose that the initial state of the system can be represented by $|0\rangle_2\equiv\tvect{0}{1}$, and the excited state by $|+\rangle_2\equiv\tvect{1}{0}$, the probability of the system occupying the $|0\rangle$ state after $N$ evolutions can be written:
\begin{equation}
P_0(N)=\left|
\begin{pmatrix}
   0  & 1
\end{pmatrix}
U^N
\begin{pmatrix}
   0 \\
   1
\end{pmatrix}
\right|^2 =  \left|\left[(\lambda^+)^N |v_0^+ |^2 + (\lambda^-)^N |v_0^- |^2\right]\right|^2,
\end{equation}
which is the absolute square of the top-left matrix element of Eq.\ (\ref{Uexpanded}). The corresponding probability of the system being in the $|+\rangle$ state is simply $P_+(N) = 1 - P_0(N)$. Since $S$ is a unitary matrix, $v_0^+$ and $v_0^-$ must satisfy $|v_0^+ |^2+|v_0^- |^2=1$, using this identity and inserting Eq.\ (\ref{lambdas}), $P_0(N)$ and $P_+(N)$ can be written:
\begin{subequations}
\begin{align}
P_0(N)&= 1-4|v_0^+ |^2|v_0^- |^2\sin^2(N[\theta_+ - \theta_-]/2)\label{pzerogeneral}\\
P_+(N)&= 4|v_0^+ |^2|v_0^- |^2\sin^2(N[\theta_+ - \theta_-]/2).\label{ponegeneral}
\end{align}
\end{subequations}
By finding $v_0^\pm$ and $\theta_\pm$ for our specific Floquet operator (\ref{floquettwostate}), we explicitly determine Eq.\ (\ref{pzerogeneral}) and (\ref{ponegeneral}), in terms of the number of pulses $N$ and the effective potential depth $V_\mathrm{eff}$, this is the origin of Eq.\ (\ref{pzero}) and (\ref{pplus}).
\subsection{Back to the system Floquet operator}
Both the amplitude $A= 4|v_0^+ |^2|v_0^- |^2$, and the oscillation frequency $\phi=\theta_+ - \theta_-$ can be determined by calculating the eigenvalues and eigenvectors of the Floquet operator (\ref{floquettwostate}), reproduced here for convenience:
\begin{equation}
F=F_{\mathrm{Free}}F_{\mathrm{Latt}}=
\begin{pmatrix}
   -1   & 0 \\
     0  &1
\end{pmatrix}
R^{-1}
\begin{pmatrix}
   e^{-2\pi i E_+ }   & 0 \\
     0  & e^{-2\pi i E_- }
\end{pmatrix}
R,
\label{floquettwostatea}
\end{equation}
where
\begin{equation}
R=
\begin{pmatrix}
   \cos(\alpha/2)   & -\sin(\alpha/2) \\
   \sin(\alpha/2)  & \cos(\alpha/2)
\end{pmatrix}.
\end{equation}
Introducing $\mu_\pm = e^{-2\pi i E_\pm}$, $\cos(\alpha/2)=c$ and $\sin(\alpha/2)=s$, we can express (\ref{floquettwostate}) in the the more compact form:
\begin{equation}
F=
\begin{pmatrix}
   -\mu_+c^2 - \mu_-s^2   & \mu_+cs - \mu_- cs \\
   -\mu_+cs + \mu_- cs  & \mu_+c^2 + \mu_-s^2
\end{pmatrix}.
\label{floquetsimplified}
\end{equation}
Using $s^2 = 1- c^2$ we can write (\ref{floquetsimplified}) as:
\begin{equation}
F=
\begin{pmatrix}
   -c^2(\mu_+ - \mu_-) - \mu_-   & cs(\mu_+ - \mu_-) \\
   -cs(\mu_+ - \mu_-)  & s^2(\mu_+ - \mu_-) + \mu_-
\end{pmatrix}.
\label{floquetsimplified2}
\end{equation}
Further, introducing the shorthand $\overline{c}^2\equiv c^2(\mu_+ - \mu_-)$, $\overline{s}^2\equiv s^2(\mu_+ - \mu_-)$, $\overline{sc}\equiv sc(\mu_+ - \mu_-)$, we have:
\begin{equation}
F=
\begin{pmatrix}
   -\overline{c}^2 - \mu_-   & \overline{sc} \\
   -\overline{sc}  & \overline{s}^2 + \mu_-
\end{pmatrix},
\label{floquetsimplified3}
\end{equation}
the eigenvalues of which can be written:
\begin{equation}
\lambda_\pm=\frac{1}{2}\left[ -\left(\overline{c}^2 -\overline{s}^2\right) \pm \sqrt{ \left(\overline{c}^2 -\overline{s}^2\right)^2 +4 \mu_- \left(\overline{c}^2 -\overline{s}^2 + \mu_- \right)} \right].
\label{lambdapm}
\end{equation}
Noting that $(\overline{c}^2 -\overline{s}^2)^2=(c^2 - s^2)(\mu_+ - \mu_-)^2$, and $(c^2-s^2)^2 = 1- 4s^2 c^2$, we can simplify the argument of the radical $(\overline{c}^2 -\overline{s}^2)^2 +4 \mu_- (\overline{c}^2 -\overline{s}^2 + \mu_-) = (\mu_+ + \mu_-)^2 - 4 s^2 c^2 (\mu_+ - \mu_-)^2$, leading to:
\begin{equation}
\lambda_\pm=\frac{(\mu_+ - \mu_-)}{2}\left[ - \left(c^2 -s^2 \right) \pm \sqrt{-4 s^2 c^2 + \left(\frac{\mu_+ + \mu_-}{\mu_+ - \mu_-} \right)^2} \right].
\label{lambdapm2}
\end{equation}
Recalling that $\mu_\pm = e^{-2\pi i E_\pm}$, and $E_{\pm} = (1 \pm \sqrt{1+ 8 V_\mathrm{eff}^2})/4$, it can be shown that 
\begin{subequations}
\begin{align}
(\mu_+ - \mu_-) &= -\left(e^{i\pi \left[E_+ - E_- \right]} - e^{-i\pi \left[E_+ - E_- \right]} \right)e^{-i\pi \left[E_+ + E_- \right]} \nonumber \\
\label{muplusminusmuminus}
&=-2\sin(\pi [E_+ - E_-]), \\ 
(\mu_+ + \mu_-) &= -\left(e^{i\pi \left[E_+ - E_- \right]} + e^{-i\pi \left[E_+ - E_- \right]} \right)e^{-i\pi \left[E_+ + E_- \right]} \nonumber \\
\label{muplusplusmuminus}
&=-2 i\cos(\pi [E_+ - E_-]), 
\end{align}
\end{subequations}
where we have made use of the fact that $E_+ + E_- = 1/2$, leading to:
\begin{equation}
\left(\frac{\mu_+ + \mu_-}{\mu_+ - \mu_-} \right)^2 = -\frac{\cos^2(\pi [E_+ - E_-])}{\sin^2(\pi [E_+ - E_-])}=-\cot^2(\pi [E_+ - E_-]).
\label{deltasquared}
\end{equation}
Since (\ref{deltasquared}) and (\ref{muplusminusmuminus}) are always real and negative, it is straightforward to separate the eigenvalues (\ref{lambdapm2}) into their real and imaginary parts:
\begin{align}
\lambda_\pm & = \mathrm{Re}(\lambda_\pm) + i\, \mathrm{Im}(\lambda_\pm) \nonumber \\ 
& =\frac{(\mu_+ - \mu_-)}{2}\left[ - \left(c^2 -s^2 \right)  \pm i \sqrt{ 4 s^2 c^2 + \delta^2 } \right],
\label{lambdapm3}
\end{align}
where we have introduced $\delta \equiv i(\mu_+ + \mu_-)/(\mu_+ - \mu_-)$ and $\delta^2 \equiv -(\mu_+ + \mu_-)^2/(\mu_+ - \mu_-)^2$. We can now solve the eigenvalue equation:
\begin{equation}
F 
\begin{pmatrix}
v_1^\pm \\
v_0^\pm
\end{pmatrix}
 =\frac{(\mu_+ - \mu_-)}{2}\left[ - \left(c^2 -s^2 \right) \pm i \sqrt{ 4 s^2 c^2 + \delta^2} \right]
\begin{pmatrix}
v_1^\pm \\
v_0^\pm
\end{pmatrix},
\label{eigenvalueeqn}
\end{equation}
for $v_0^\pm$, $v_1^\pm$. Equation (\ref{eigenvalueeqn}) leads directly to:
\begin{equation}
v_1^\pm = i \left( \epsilon \pm \sqrt{\epsilon^2 + 1} \right)v_0^\pm,
\end{equation}
where we have introduced the shorthand $\epsilon \equiv -\delta/2sc$. We can now state that:
\begin{equation}
\vec{v}_+ \propto  \left( \begin{array}{c}
i \left[ \epsilon + \sqrt{\epsilon^2 + 1} \right] \\
1
\end{array}
\right),
\quad
\vec{v}_- \propto \left( \begin{array}{c}
i \left[ \epsilon - \sqrt{\epsilon^2 + 1} \right] \\
1
\end{array}
\right),
\end{equation}
and noting that $\sqrt{\epsilon^2 +1} - \epsilon = (\sqrt{\epsilon^2 +1} +\epsilon )^{-1}$, we can express the normalized eigenvectors thus:
\begin{subequations}
\begin{align}
\vec{v}_+ &= \frac{1}{\sqrt{2 \sqrt{\epsilon^2 + 1}}} \left( \begin{array}{c}
i \sqrt{\sqrt{\epsilon^2 + 1}+ \epsilon} \\
\sqrt{\sqrt{\epsilon^2 + 1}- \epsilon}
\end{array}
\right), \label{vplusvec} \\
\vec{v}_- &= \frac{1}{\sqrt{2 \sqrt{\epsilon^2 + 1}}} \left( \begin{array}{c}
\sqrt{\sqrt{\epsilon^2 + 1}- \epsilon} \\
i \sqrt{\sqrt{\epsilon^2 + 1}+ \epsilon}
\end{array}
\right). \label{vminusvec}
\end{align}
\end{subequations}
The amplitude A=$4|v_0^+ |^2|v_0^- |^2$ can now be determined from the product of the absolute squares of the bottom entries of $\vec{v}_+$ and $\vec{v}_-$:
\begin{align}
A &= \frac{4}{\left( \sqrt{2 \sqrt{\epsilon^2 + 1}} \right)^4} \left[ \left(\sqrt{\epsilon^2 + 1} - \epsilon \right)\left(\sqrt{\epsilon^2 + 1} + \epsilon \right) \right] \nonumber \\
 &= \frac{1}{\epsilon^2 +1}.
\end{align}
Inserting $\epsilon^2 = \delta^2/4s^2c^2$ and $4s^2c^2=\sin^2(\alpha)=\sin^2(\arcsin( 2\sqrt{2} V_\mathrm{eff}/\sqrt{1 + 8 V_\mathrm{eff}^2} )) = 8V_\mathrm{eff}^2/(1 + 8V_\mathrm{eff}^2)$ we can express the amplitude in terms of the effective lattice-depth $V_\mathrm{eff}$:
\begin{equation}
A = \frac{8 V_\mathrm{eff}^2 \sin^2 \left(\pi \sqrt{1+8V_\mathrm{eff}^2}/2\right)}{8 V_\mathrm{eff}^2 + \cos^2\left(\pi \sqrt{1+8V_\mathrm{eff}^2}/2\right)},
\label{amplitude}
\end{equation}
which corresponds to Eq.\ (\ref{amplitudea}). Using Eq.\ (\ref{lambdapm3}), we can also determine the oscillation frequency $\phi=\theta_+ - \theta_- = \mathrm{arg}(\lambda_-) - \mathrm{arg}(\lambda_+)$. We can express $\phi$ as:
\begin{align}
\phi &=\arctan\left(\frac{\mathrm{Im}(\lambda_-)}{\mathrm{Re}(\lambda_-)}\right) - \arctan\left(\frac{\mathrm{Im}(\lambda_+)}{\mathrm{Re}(\lambda_+)}\right) \nonumber \\
& = 2\, \arctan\left(\frac{\mathrm{Im}(\lambda_-)}{\mathrm{Re}(\lambda_-)}\right),
\end{align}
where we have used the relations $\mathrm{Re}(\lambda_-)=\mathrm{Re}(\lambda_+)$, and $\mathrm{Im}(\lambda_+)=-\mathrm{Im}(\lambda_-)$. Substituting in $\mathrm{Re}(\lambda_-)=-(\mu_+ - \mu_-)(c^2-s^2)/2$ and $\mathrm{Im}(\lambda_-)=-(\mu_+ - \mu_-)\sqrt{4s^2c^2 + \delta^2}/2$ we have:
\begin{equation}
\phi=2\,\arctan\left(\frac{\sqrt{4s^2c^2 + \delta^2}}{c^2-s^2}\right),
\end{equation}
which, noting that $4s^2c^2=8V_\mathrm{eff}^2/(1+8V_\mathrm{eff}^2)$ and recalling that $\delta^2=\cot^2(\pi \sqrt{1+8V_\mathrm{eff}^2}/2)$, can be written:
\begin{equation}
\nonumber
\phi = 2\,\arctan\left(\frac{\sqrt{8V_\mathrm{eff}^2 + \cos^2\left(\pi \sqrt{1+8V_\mathrm{eff}^2}/2\right)}}{\sin\left(\pi \sqrt{1+8V_\mathrm{eff}^2}/2\right)}\right),
\end{equation}
which corresponds to Eq.\ (\ref{frequencyphi}).
\section{Limiting behaviours of Equations (\ref{amplitudea}) and (\ref{frequencyphi}) \label{limitsappendix}}
\subsection{Weak coupling regime, $\boldsymbol{V_\mathrm{eff}\rightarrow 0}$}
\label{appendixweakcoupling}
Equation (\ref{frequencyphi}) can be linearized in the weak coupling regime as $V_\mathrm{eff}\rightarrow 0$. To clarify the procedure, we introduce the following notation:
\begin{subequations}
\begin{align} 
\label{phixy}
\phi &=2\arctan\left(\frac{Y}{X}\right), \\
\label{phix}
Y &= \sqrt{8V_\mathrm{eff}^2 + \cos^2\left(\pi\sqrt{1+8V_\mathrm{eff}^2}/2\right)}, \\
\label{phiy}
X &= \sin\left(\pi\sqrt{1+8V_\mathrm{eff}^2}/2\right).
\end{align}
\end{subequations}
Clearly as $V_\mathrm{eff}\rightarrow 0$, it follows that $Y \rightarrow \cos(\pi/2) = 0$, $X \rightarrow \sin(\pi/2) = 1$, and therefore $\phi \rightarrow 2 \arctan(0/1) = 0$. However, we can still find an approximation to $\phi$ that is linear in $V_\mathrm{eff}$ by means of a Taylor expansion:
\begin{equation}
\phi=2\arctan(Z)\approx Z - \frac{Z^3}{3}+\frac{Z^5}{5}\dots,
\end{equation}
where $Z=Y/X$. Hence, near $V_\mathrm{eff}=0$, $\phi$ is given approximately by $\phi\approx 2Y/X$. Note that $\sin(\theta)= \cos(\theta - \frac{\pi}{2})$, $\cos(\theta)= -\sin(\theta - \frac{\pi}{2})$,
and hence
\begin{subequations}
\begin{align}
\label{shiftsin}
\sin\left(\pi\sqrt{1+8V_\mathrm{eff}}/2\right)&= \cos\left(\pi\left[\sqrt{1+8V_\mathrm{eff}}-1\right]/2\right),\\
\label{shiftcos}
\cos\left(\pi\sqrt{1+8V_\mathrm{eff}}/2\right)&= -\sin\left(\pi\left[\sqrt{1+8V_\mathrm{eff}}-1\right]/2\right).
\end{align}
\end{subequations}
The arguments of the trigonometric functions on the right hand side tend to zero as $V_\mathrm{eff} \rightarrow 0$, which simplifies the expansions of (\ref{shiftsin}) and (\ref{shiftcos}), since we can use standard small-angle approximations. We can simplify the arguments further by use of the binomial approximation $\sqrt{1+\epsilon}\approx 1 + \epsilon/2$, yielding:
\begin{subequations}
\begin{align}
\cos\left(\pi\left[\sqrt{1+8V_\mathrm{eff}}-1\right]/2\right) &\approx \cos(2\pi V_\mathrm{eff}^2)\approx 1-\frac{4\pi^2 V_\mathrm{eff}^4}{2},\\
\sin\left(\pi\left[\sqrt{1+8V_\mathrm{eff}}-1\right]/2\right) &\approx \sin(2\pi V_\mathrm{eff}^2)\approx 2\pi V_\mathrm{eff}^2.
\end{align}
\end{subequations}
Hence, carrying out these approximations subsequent to substituting Eq.\ (\ref{shiftsin}) into Eq.\ (\ref{phix}) and Eq.\ (\ref{shiftcos}) into Eq.\ (\ref{phiy}):
\begin{align}
Y &= \sqrt{8V_\mathrm{eff}^2 + \sin^2\left(\pi\left[\sqrt{1+8V_\mathrm{eff}^2} -1 \right]/2\right)} \nonumber\\
&\approx\sqrt{8V_\mathrm{eff}^2 + 4\pi^2 V_\mathrm{eff}^4} 
\approx 2\sqrt{2} V_\mathrm{eff},\\
X &= \cos\left(\pi \left[ \sqrt{1+8V_\mathrm{eff}^2} -1 \right]/2 \right) \nonumber \\
& \approx \cos\left( 2\pi V_\mathrm{eff}^2 \right) \approx 1 - 2 \pi^2 V_\mathrm{eff}^4 \approx 1.
\end{align}
Therefore, to leading order in $V_\mathrm{eff}$, around $V_\mathrm{eff}=0$,
\begin{equation}
\label{phiapproximant}
\phi \approx \frac{2\times 2\sqrt{2} V_\mathrm{eff}}{1} = 4\sqrt{2}V_\mathrm{eff}.
\end{equation}
We may follow a similar procedure for Eq.\ (\ref{amplitudea}), reproduced here for convenience:
\begin{equation}
A = \frac{8 V_\mathrm{eff}^2 \sin^2\left(\pi \sqrt{1+8V_\mathrm{eff}^2}/2\right)}{8 V_\mathrm{eff}^2 + \cos^2\left(\pi \sqrt{1+8V_\mathrm{eff}^2}/2\right)}.
\end{equation}
Using Eqs.\ (\ref{shiftsin}) and (\ref{shiftcos}), it follows that, around $V_\mathrm{eff}=0$, $\sin^2(\pi\sqrt{1+8V_\mathrm{eff}^2}/2) \approx 1$ and $\cos^2(\pi\sqrt{1+8V_\mathrm{eff}^2}/2) \approx 0$,
leading to:
\begin{equation}
\label{aapproximant}
A\approx\frac{8V_\mathrm{eff}^2\times1}{8V_\mathrm{eff}^2+0}\approx1.
\end{equation}
\subsection{Strong coupling regime, $\boldsymbol{V_\mathrm{eff}\rightarrow \infty}$}
To determine the behavior of $\phi$ as $V_\mathrm{eff}\rightarrow \infty$ we first rearrange Eq.\ (\ref{phix}):
\begin{align}
Y&=\sqrt{8V_\mathrm{eff}^2 + \cos^2\left(\pi\sqrt{1+8V_\mathrm{eff}^2}/2 \right)} \nonumber \\
&= 2\sqrt{2} V_\mathrm{eff} \left[ 1 + \frac{\cos^2\left(\pi\sqrt{1+8V_\mathrm{eff}^2}/2 \right)}{16V_\mathrm{eff}^2}\right].
\end{align}
Clearly, as $V_\mathrm{eff}\protect{\rightarrow} \infty$, $Y\protect{\approx} 2\sqrt{2} V_\mathrm{eff}$, whereas $X\protect{=}\sin( \pi \sqrt{1+8V_\mathrm{eff}^2 / 2} )$ simply oscillates. Therefore, recalling Eq.\ (\ref{phixy}), if $X=0$ and $Y>0$, then $\phi=\pi$. Also, for nonzero $X$, then as $V_\mathrm{eff}\rightarrow\infty$, $Y\rightarrow\infty$, and therefore $\phi\rightarrow\pi$, either from below ($X>0$) or above ($X<0$). The curve of $\phi$ as a function of $V_\mathrm{eff}$ crosses through the line where $\phi = \pi$ whenever $\pi\sqrt{1+8V_\mathrm{eff}^2}=m\pi$ for $m\in\mathbb{Z}^+$, in other words where: 
\begin{equation}
V_\mathrm{eff}=\sqrt{\frac{4m^2-1}{8}},
\end{equation}
or, as $V_\mathrm{eff}\rightarrow\infty$,
\begin{equation}
V_\mathrm{eff}=\frac{m}{\sqrt{2}}.
\end{equation}
\subsection{Quadratic approximant to Equation (\ref{pplus})}
\label{heroldlimit}
Equation (\ref{pplus}) can be rewritten in terms of the first few orders of a Taylor expansion:
\begin{equation}
P_+(N,V_\mathrm{eff})=A\sin^2(x)\approx Ax^2 - \frac{A}{3}x^4+\cdots,
\end{equation}
with $x\equiv N\phi/2$, in a regime where $x\ll1$. Further, assuming that $V_\mathrm{eff}$ is near zero, we may replace $\phi$ and $A$ with our leading  order approximations of Eqs.\ (\ref{phiapproximant},\ref{aapproximant}), with $x\approx2\sqrt{2}NV_\mathrm{eff}$. Hence, to leading (quadratic) order in $x$:
\begin{equation}
P_+(N,V_\mathrm{eff})\approx 8N^2V_\mathrm{eff}^2\protect{\propto}N^2,
\end{equation}
which corresponds to the result used in \cite{herold_matrix_elements,kao_tuneout_dysprosium} where $P_+ \equiv P_1$ and $V_\mathrm{eff}=V_0/(16E_\mathrm{R})=U_0/(16E_\mathrm{R})$.

\section{Numerical diagonalization \label{numerical_diagonalization}}
To diagonalize the lattice Hamiltonian in the zero-quasimomentum subspace, we first express Eq.\ (\ref{Hlatt}) as:
\begin{equation}
\frac{M}{\hbar^2 K^2} \hat{H}_{\mathrm{latt}} = \tilde{H}_{\mathrm{latt}} = \frac{\hat{k}^2}{2} - \frac{V_\mathrm{eff}}{2} \left(e^{i 2k_l \hat{x}} + e^{-i 2k_l \hat{x}}\right).
\label{HlattReduced}
\end{equation}
Here $e^{i 2k_l \hat{x}}$ and $e^{-i 2k_l \hat{x}}$ are momentum displacement operators, which act on the momentum eigenkets in the following way:
\begin{align}
e^{i 2k_l \hat{x}}| k=\alpha \rangle &= | k=\alpha + 1 \rangle,
& e^{-i 2k_l \hat{x}}| k=\alpha \rangle &=| k=\alpha - 1 \rangle.
\end{align}
The matrix elements of the Hamiltonian can, therefore, be expressed in the momentum basis thus:
\begin{align}
\tilde{H}_\mathrm{latt\, \gamma,\alpha}&=\langle k=\alpha |\tilde{H}_\mathrm{latt}| k=\gamma \rangle 
= \frac{\gamma^2}{2}\delta_{\gamma,\alpha} - \frac{V_\mathrm{eff}}{2}(\delta_{\gamma,\alpha-1} + \delta_{\gamma,\alpha+1} ) \nonumber \\
&=\frac{\gamma^2}{4}\delta_{\gamma,\alpha} - \frac{V_\mathrm{eff}}{2}\delta_{\gamma,\alpha-1} + \mathrm{H.c.},
\label{HlattElements}
\end{align}
where $\alpha,\gamma \in \mathbb{Z}$. Equation (\ref{HlattElements}) can then be expressed in matrix form, and numerically diagonalized in order to find the time evolution of an initial momentum eigenstate.

By expressing Eq. \ref{HlattElements} in matrix form thus:
\begin{equation}
H_{\mathrm{latt}}=
\begin{pmatrix}
    \ddots & \vdots & \vdots & \vdots  &\reflectbox{$\ddots$} \\
   \dots & 1/2   & -V_{\mathrm{eff}}/2 & 0  & \dots  \\
  \dots  &-V_{\mathrm{eff}}/2  & 0 & -V_{\mathrm{eff}}/2 & \dots \\
  \dots  &0   &- V_{\mathrm{eff}}/2 & 1/2 & \dots  \\
   \reflectbox{$\ddots$} & \vdots & \vdots & \vdots & \ddots  \\
\end{pmatrix},
\end{equation}
we can construct the matrix $P^{n\times n}$ diagonalizing $H_{\mathrm{latt}}^{n\times n}$, such that $H_\mathrm{latt, diag}^{n\times n} = (P^{\dagger})^{n\times n} H_{\mathrm{latt}}^{n\times n}P^{n\times n}$.
We are led to the expression:
\begin{equation}
|\psi(t=N) \rangle^{n \times 1} =[ H_\mathrm{free}^{n\times n}P^{n\times n} H_\mathrm{latt, diag}^{n\times n}(P^{n\times n})^\dagger]^N | K=\alpha \rangle^{n \times 1},
\end{equation}
for $|\psi(t=N) \rangle^{n \times 1}$, the time evolution due to $N$ pulse sequences of an initial eigenstate $| K=\alpha \rangle^{n \times 1}$, where $\alpha \in [-(n-1)/2,(n-1)/2]$. The $n\times1$ superscript denotes that the ket should be understood as an $n$-dimensional column vector.

\end{document}